# A Size-Consistent Wavefunction Ansatz Built from Statistical Analysis of Orbital Occupations

Valerii Chuiko[1], Paul W. Ayers[1]

[1]McMaster University, Hamilton, Canada

Direct approaches to the quantum many-body problem suffer from the so-called "curse of dimensionality": the number of parameters needed to fully specify the exact wavefunction grows exponentially with increasing system size. This motivates the develop of accurate, but approximate, ways to parameterize the wavefunction, including methods like coupled cluster theory and correlator product states (CPS). Recently, there has been interest in approaches based on machine learning, both direct application of neural network architecture and the combination of conventional wavefunction parameterizations with various Boltzmann machines. While all these methods can be exact in principle, they are usually applied with only a polynomial number of parameters, limiting their applicability.

This research's objective is to present a fresh approach to wavefunction parameterization that is size-consistent, rapidly convergent, and robust numerically. Specifically, we propose a hierarchical wavefunction ansatz that converges rapidly (with respect to the number of orbital-couplings included) and which is computationally robust (as it is based on least-squares optimization). The general utility of this approach is verified by applying it to uncorrelated, weakly-correlated, and strongly-correlated systems, including small molecules and the one-dimensional Hubbard model.

# Table of Contents

# Introduction

The study of chemical systems at a quantum level is an essential aspect of many scientific disciplines, including materials science[1–3], pharmaceuticals[4–6], and catalysis[7,8]. The understanding of the properties of molecules, such as their reactivity and stability, is fundamental to the design and development of new materials, drugs, and catalysts. The wave function, which describes the quantum state of a molecule, is a central concept in theoretical and computational chemistry. It provides a complete description of the electronic structure of a molecule, allowing researchers to predict a wide range of chemical properties, such as molecular energies, electronic states, and reaction rates.

Despite its importance, exact solutions to the Schrödinger equation, which describes the behavior of quantum systems, are often difficult or impossible to obtain for complex systems. Traditional wave function methods rely on approximations and assumptions that limit their accuracy and applicability. Let's look at one of the most widely used methods to solve the Schrödinger equation.

Full Configuration Interaction (FCI) is a powerful method for obtaining the exact wave function of a quantum mechanical system. The method involves calculating the wave function as a linear combination of all possible configurations of the system's electrons. This means that the FCI method can provide highly accurate results that can be used to study complex quantum systems. The central idea of this method is to rewrite $N-$ electron basis function

One of the advantages of FCI is its ability to provide the exact wave function $|\Phi_i\rangle$ as substitution from the Hartree-Fock reference determinant:

$$|\Psi\rangle = c_0|\Phi_0\rangle + \sum_{ra} c_a^r|\Phi_a^r\rangle + \sum_{a<b,r<s} c_{ab}^{rs}|\Phi_{ab}^{rs}\rangle + \sum_{r<s<t,a<b<c} c_{abc}^{rst}|\Phi_{abc}^{rst}\rangle + \cdots$$

where $|\Phi_a^r\rangle$ means the Slater determinant formed by replacing spin-orbital $a$ in $|\Phi_0\rangle$ with spin orbital $r$, etc. Every $N-$ electron Slater determinant can be described by the set of $N$ spin orbitals from which it is formed, and this set of orbital occupancies is often referred to as a "configuration." This is particularly useful in studying complex quantum systems where approximate methods may not provide accurate results. FCI has been used to study a wide variety of quantum systems, from small molecules to large solids. In particular, FCI is critical for understanding and predicting chemical reactivity by accurately determining the electronic structure of molecules.

However, FCI has a major limitation in that its computational cost increases exponentially with the number of electrons and orbitals. This makes it infeasible to apply FCI to larger systems. To overcome this limitation, several approaches have been proposed to reduce the computational cost of FCI.

One approach is the Davidson diagonalization method, which can be used to obtain only the most significant configurations of the wave function. This reduces the computational cost of FCI significantly while maintaining a high degree of accuracy. Another approach is the direct CI method, which uses a truncated Hilbert space to reduce the number of configurations included in the wave function. However, these methods still suffer from computational limitations when applied to larger systems. Specifically, despite a lot of optimization parameters that grows exponentially with increasing size of the system, many CI coefficients are very small and almost don't significantly contribute to the total energy[9–11].

Over the years, several methods have been developed to tackle this problem, including coupled-cluster theory[12–16], correlator product state (CPS)[17,18], Monte Carlo sampling technics, density matrix renormalization group (DMRG)[19–22], family of methods based on neural networks and restricted Boltzmann machines[23–29]. However, these methods are computationally expensive and not scalable to large systems.

For example the Correlator product state method[17]. This method aims to represent the wave function of a many-body system by utilizing a product of local correlation functions. This method has shown to be efficient in representing the wave function and has been applied to various systems. CPS are obtained by associating variational degrees of freedom directly with correlations between groups of sites. For example, in the nearest-neighbor two-site CPS, a correlator is associated with each neighboring pair of site:

$$|\Psi\rangle = \sum_{\{q\}} \prod_{\langle ij \rangle} C^{q_i q_j} |q_1, \ldots, q_L\rangle$$

here $< ij >$ denotes the nearest neighbours.

The method is particularly suitable for systems with a small number of strongly correlated electrons, such as transition metal complexes. However, even though these approach scales polynomial with the increasing number of electrons, systematic improving its performance is not a straightforward task[30].

Recently, a new approach to reduce the computational cost of the FCI method has been proposed, called the adaptive sampling algorithm. This method involves selecting a subset of the most important configurations of the wave function using a sampling technique. This family of based is based on Monte Carlo sampling[10,11], which involves generating random samples from a probability distribution to approximate the wave function.

The basic idea is to rewrite energy of the system in the ground state as:

$$E = \frac{\int d\mathbf{r} \psi^*(\mathbf{r}) \hat{H} \psi(\mathbf{r})}{\int d\mathbf{r} \psi^*(\mathbf{r}) \psi(\mathbf{r})} = \frac{\int d\mathbf{r} \psi^*(\mathbf{r}) \psi(\mathbf{r}) \frac{1}{\psi(\mathbf{r})} \hat{H} \psi(\mathbf{r})}{\int d\mathbf{r} \psi^*(\mathbf{r}) \psi(\mathbf{r})}$$

$$= \int d\mathbf{r} \frac{|\psi(\mathbf{r})|^2}{\int d\mathbf{r} |\psi(\mathbf{r})|^2} \frac{1}{\psi(\mathbf{r})} \hat{H} \psi(\mathbf{r}) = \int d\mathbf{r} \rho(\mathbf{r}) E_l(\mathbf{r})$$

where $\psi(\mathbf{r})$ is the wave function of the system, $\hat{H}$ is the Hamiltonian operator, and $\mathbf{r}$ represents the positions of all the particles in the system, $\rho(\mathbf{r}) = \frac{|\psi(\mathbf{r})|^2}{\int d\mathbf{r} |\psi(\mathbf{r})|^2}$, $E_l(\mathbf{r}) = \frac{1}{\psi(\mathbf{r})} \hat{H} \psi(\mathbf{r})$. At each step N generates $\mathbf{r}_N$ and uniform number $u$. If $u \leq \frac{\rho(\mathbf{r}_N)}{\rho(\mathbf{r}_{N-1})}$ then accept $\mathbf{r}_N$ else reject $\mathbf{r}_N$ and set $\mathbf{r}_N = \mathbf{r}_{N-1}$. Such iterative process continues till algorithm is converged. Monte Carlo sampling methods have been shown to be effective for systems with large numbers of particles, where other methods may not be feasible. That is why they widely used to speed up the computations for the various ansatz.

Another approach utilizes neural networks and restricted Boltzmann machines (RBM)[23–29], which have emerged as a promising approach to capture correlation energy in quantum mechanics simulations. Neural networks can learn complex non-linear relationships between inputs and outputs, making them suitable for various systems. The method has been applied to the study of chemical reactions, metal-organic frameworks, and materials discovery.

The main problem is this family of methods requires enormous amount of training data and they can predict properties only of similar systems such as different geometries of the same molecule which makes them almost not applicable in the real-world scenarios.

      Overall, these recent approaches provide a promising direction for accurately capturing correlation energy in quantum mechanics simulations. However, further research is needed to explore the applicability of these methods to larger and more complex systems. Moreover, the combination of these methods with the traditional many-body techniques may provide new tools for the accurate description of correlation energy in quantum mechanics simulations.

## Method

In light of the shortcomings of existing methods, the objective of this paper is to present a fresh approach to the wavefunction approximation that meets the following key criteria:

1. Size Consistency: The proposed wavefunction must exhibit size consistency, meaning that the results of the calculation should not qualitatively depend on the size of the system being studied.
2. Fast Convergence: To ensure efficiency and practicality, the method must converge quickly, producing results in a timely manner.
3. Robust Computational Performance: The numerical algorithm used to implement the wavefunction must be computationally robust, meaning that it should be able to handle a wide range of inputs and produce reliable results even in the presence of challenging conditions such as presence of highly correlated systems.

By fulfilling these criteria, the novel wavefunction ansatz introduced in this paper aims to overcome the limitations of existing methods and provide a new and improved approach to this field of research.

The goal is to simplify the Configuration Interaction (CI) coefficient by expressing it as a function of only a few occupation numbers. This might be possible because, in general, each orbital is strongly entangled with only a few other orbitals.

According to the proposed criteria that should be met, the best way to represent wavefunction is using the second quantization. Specifically, Fock space provides several advantages:

1. Abstraction of Quantum States: Fock space provides a mathematical framework to abstractly describe quantum states as a linear combination of particle creation and annihilation operators, instead of wave functions.
2. Simplification of Interactions: Fock space also makes it easier to describe the interactions between particles, as they can be described as a sum of simple pairwise interactions.
3. Extension to Many-Body Systems: Fock space can be extended to describe many-body systems, making it useful for studying condensed matter physics, quantum chemistry and solid-state physics.

Overall, Fock space provides a more intuitive and computationally tractable representation of quantum systems, which is particularly useful for studying many-body systems in highly correlated regime.

The aim is to write the CI coefficient as a function of K spin-orbitals $f(n_1, n_2, \ldots, n_K)$. The underlying assumption is that the probability of finding an electron in a particular orbital can be used to determine the relationship between the spin-orbitals and the CI coefficient.

Summing up all the criteria mentioned above, we can assume that the probability of finding an electron, which is associated with the wavefunction represented by the occupation number vector $\vec{n}$, can be expressed as follows:

$$c_{\vec{n}} = e^A \prod_i e^{w n_i} \cdot \prod_{ij} e^{w_{ij} n_i n_j} \prod_{ijk} e^{w_{ijk} n_i n_j n_k} \ldots \quad (1)$$

here $c_{\vec{n}}$ is the coefficient obtained using Full CI method that corresponds to the determinant with occupation number vector $\vec{n}$; $n_i = 1$ if $i-th$ orbital is occupied in the corresponding determinant and 0 if it is not; $A, \omega_I, \omega_{ij}, \omega_{ijk} \ldots$ – some constants that need to be optimized. Such decomposition should be size consistent given its multiplicative form. However, such function is hard to optimize because of the exponential form, but corresponding minimization problem is logarithmically convex and can be simplified to a regular least squared problem:

$$\ln c_{\vec{n}} = A + \sum_i w_i n_i + \sum_{ij} w n_i n_j + \sum_{ijk} w_{ijk} n_i n_j n_k + \cdots \quad (2)$$

Here we use the complex logarithm, allowing the coefficients of regression to be complex. After taking the logarithm, the LHS of the equation (2) yields a phase that denotes the sign of the initial CI coefficient:

$$\ln|c_{\vec{n}}| + k\pi i = \ln c_{\vec{n}},$$

where k is even if $c_{\vec{n}} > 0$ and odd otherwise.

Our goal is to find such coefficients $\omega$ so that equation (1) is satisfied. In this paper we propose to fit the absolute values of coefficients first, and then fit the complex part so that:

$$w = \omega + vi,$$

here $v$ is either 0 or $\pi$.

## Regression of the absolute value

Because $0 < |c_{\vec{n}}| \leq 1$ the LHS is always negative, it makes sense rewrite the following expansion as

$$-\ln|c_{\vec{n}}| = A + \sum_i \omega_i n_i + \sum_{ij} \omega_i n_i n_j + \sum_{ijk} \omega_{ijk} n_i n_j n_k + \cdots$$

With the constraint, $\omega_i, \omega_{ij}, \omega_{ijk} \ldots \geq 0$. Applying this constraint we can interpret $\omega_i, \omega_{ij}, \omega_{ijk}$ as probability to find electrons $i, (i,j), (i,j,k)$ occupying spin orbitals $i, (i,j), (i,j,k)$ respectively; also, nonnegative optimization solves the problem with ill-conditioned feature vector that is represented by combinations of occupations of orbitals.

So, the corresponding minimization problem is:

$$\min_{\substack{\{\omega_i, \omega_{ij}, \omega_{ijk} \ldots\} \\ \omega_i \geq 0 \\ \omega_{ij} \geq 0 \\ \omega_{ijk} \geq 0}} \sum_n \left( -\ln|c_{\vec{n}}| - A + \sum_i \omega_i n_i + \sum_{ij} \omega_i n_i n_j + \sum_{ijk} \omega_{ijk} n_i n_j n_k + \cdots \right)^2 \quad (2)$$

Such minimization isn't trivial considering the that $-\ln|c_{\vec{n}}|$ range from 0 to $\infty$ for big CI coefficients (~1) and small (~$10^{-10}$) which means tends to assign large weights to zero coefficients and small weights to non-zero coefficients, leading to inaccurate results.

To address this issue, several standard approaches have been tried, including weighted linear regression and Iteratively Reweighted Linear Regression. The most trivial choices for weights are absolute or square value of the CI coefficient: $abs(c_{\vec{n}}), c_{\vec{n}}^2$. Such choice of weights perfectly fits the big coefficients but does not consider zero coefficients, so their predicted values given the optimized have the biggest error. These approaches qualitatively are the same as fitting only non-zero coefficients which reduces the accuracy of the model. Another way to address this problem is to apply Iteratively Reweighted Linear Regression. Weights are updated based on the error of the previous step such as:

$w_i^{(t)} = |y_i - X_i \beta^{(t)}|^{p-2}$, $w_i^{(t)} = \frac{1}{max\{\delta, |y_i - X_i \beta^{(t)}|\}}$, $w_i = \frac{X_i \beta}{\ln \frac{y_i}{X_i \beta}}$; where $y_i = -\ln c_i^2$, $X_i - i$-th row of the feature matrix, $\beta^{(t)}$ – vector of regression coefficients obtained at the step t, $\delta, p$ – hyperparameters. However, these reweighed technics don't work for proposed decomposition and method don't converge. The last choice was inspired by KL Divergence metric[31]. Another iterative method that we tried is iterative ordinary least square (iOLS)[32], specifically the transformation for the Poisson regression case:

$$\tilde{Y}_i(\beta, \delta) = \log(Y_i + \delta \exp(X_i' \beta)) - c(\beta, \delta),$$

$$c_i(\delta, \beta) = \log\left(\delta + Y_i \exp(-X_i' \beta)\right) - \frac{1}{1+\delta}(Y_i \exp(-X_i' \beta) - 1)$$

Here $\delta$ is hyper parameter. The proposed model converges, but with poor fitting of the significant ($> 0.5$) CI coefficients which may be a result of many zeros.

Another problem for this specific minimization problem in is that number of zero coefficients are much bigger than non-zero coefficients and the values of them (zero coefficients) can be represented numerically in many ways e.g. $10^{-10}, 10^{-20}, 10^{-8}$ etc. but have to be treated equally. Zero Inflated Poisson Model[33–35] is widely used in such cases. The assumption is that CI coefficients are generated by two random processes:

$y_i \sim 0$ with probability $p_i$
$\sim$ Poisson $(\lambda_i)$ with probability $1 - p_i$,

So that:

$y_i = 0$ with probability $p_i + (1 - p_i)e^{-\lambda_i}$
$= k$ with probability $(1 - p_i)e^{-\lambda_i}\lambda_i^k/k!$,
$k = 1, 2, ...$

This method also showed a poor performance, because it assumes that there is enough data for fitting Poisson Distribution and separate it from the zero values, generated by Bernoulli process. However, there is not sufficient non-zero coefficients to fit the "Poisson" part.

The proposed solution to the minimization problem of coefficients follows a two-step procedure. First, the Maximum likelihood-based parameter estimation is applied to fit the problem to a Poisson distribution by maximizing the likelihood function:

$$\ell(\beta \mid X, Y) = \log L(\beta \mid X, Y) = \sum_{i=1}^{m} \left(y_i \beta x_i - e^{\beta x_i} - \log(y_i!)\right)$$

Here $L(\beta \mid X, Y) = \prod_{i=1}^{m} \frac{e^{y_i \beta x_i} e^{-e^{\beta x_i}}}{y_i!}$ – likelihood function.

Maximization of $\ell(\beta \mid X, Y)$ is equivalent to minimization $-\ell(\beta \mid X, Y)$. So the initial optimization problem can be rewritten as:

$$\min_{\omega_i, \omega_{ij}, ... \geq 0} \sum_{i=1}^{m} (e^{-\omega x_i} + y_i \omega x_i) \qquad (3)$$

This is a convex function and for the minimization algorithm was chosen L-BFGS. The second step involves refining the set of $\boldsymbol{\omega}$ values obtained in the first step by optimizing the sum of squared differences between the target coefficients and a product of exponential functions:

$$\min_{\omega_i, \omega_{ij}, \ldots \geq 0} \sum_n \left( c_{\vec{n}} - e^{-A} \prod_i e^{-\omega_i n_i} \cdot \prod_{ij} e^{-\omega_{ij} n_i n_j} \prod_{ijk} e^{-\omega_{ijk} n_i n_j n_k} \ldots \right)^2 \quad (4)$$

For the optimization algorithm is chosen "truncated newton" from the scipy.optimize.minimize implementation.

## Regression of the complex phase

In order to learn the sign of the predicted value of CI coefficient we propose to use weighted nonnegative least square minimization algorithm for the complex part of the ansatz. We can do this using the same feature matrix $X$ that used to predict the square values of CI coefficients. So, the objective function to be minimized is:

$$\min_{\substack{\{v, v_{ij}, v_{ijk} \ldots\} \\ v_i \geq 0 \\ v_{ij} \geq 0 \\ v_{ijk} \geq 0}} \sum_n \left( |c_{\vec{n}}| \left( \phi + A + \sum_i v n_i + \sum_{ij} v_i n_i n_j + \sum_{ijk} v_{ijk} n_i n_j n_k + \cdots \right) \right)^2 \quad (5)$$

Here $\phi$ is the phase that is equal to either 0 or $\pi$ for the positive and negative CI coefficients respectively.

Such minimization problem is the same as described at equation (2), but it doesn't have the same problem – even if the predicted sign of the coefficient is wrong it doesn't matter, because its value is close to zero.

For the minimization technique we use standard nonnegative least square minimization algorithm implemented in scipy.optimize.nnls. When the weight $v$ are calculated, regressed coefficients can be any number from $(-\infty; +\infty)$, so we induced periodicity of the phase, by calculating the ceil division of the regressed coefficient and rounding them to $\pi$ if $\frac{\pi}{2} < \phi_{predicted} < \frac{3\pi}{2}$ and to 1 otherwise.

It worth noting that interaction of $i$-th and $j$-th fermions in the state represented by product of occupation numbers $n_i n_j$ defines state of the system more precisely than linear combination of the occupation numbers of the same electrons in states $n_i, n_j$ respectively. From now on we define the $k_{th}$ order of approximation of wavefunction that includes only set of features:

$$X^k = \prod_{i=1}^{K} \prod_{j=1}^{N_{C_k}} n_{i_1} n_{i_2} \ldots n_{i_j} \quad (6)$$

Here $N_{C_k}$ – number of combinations $C_K^k = \frac{K!}{k!(K-k)!}$; $K$ –number of spin-orbitals of the system.

## Modeled systems

To validate the proposed method, systems with a range of sizes, ranging from 4 to 14 electrons, and covering both weak and strong correlations were selected for testing. Specifically we modeled different systems that described by spinless Hubbard model at half-filing:

$$\hat{H}_{\text{Hubbard}} = \sum_{pq} h_{pq} a_p^\dagger a_q + \sum_p U_p \hat{n}_{p\alpha} \hat{n}_{p\beta}$$

Where one body term defined using Huckel model:

$$h_{pq} = \begin{cases} \alpha \text{ if } p = q \\ \beta \text{ if } p \text{ is connected to } q \\ 0 \text{ otherwise} \end{cases}$$

Traditionally, the values of $\alpha$ and $\beta$ are defined for the 2p orbital in an sp2 hybridized carbon atom, with reasonable values being: $\alpha = -11.26 \, eV = 0.414$ Hartree; $\beta = -1.45 \, eV = 0.0533$ Hartree. The $U_p$ term denotes the repulsion of electrons on the same atom/group site. However, to be consistent with[17,28,29] we used $\alpha = 0, \beta = -1$ Hartree.

In the current study, the 6 and 10 site Hubbard model was used to test the proposed ansatz. To model these systems, a natural orbital basis was used, which allows for a more efficient and accurate representation of the electronic wavefunction along with adequate treatment of dynamic correlations[36].

The main idea is to present an orbital as a eigen vector of the first-order reduced density matrix, obtained from an N-electron wave function $\Psi$:

$$\gamma(1' \mid 1) = N \int \Psi(1', 2, \ldots N) \Psi^*(1, 2, \ldots N) d\tau_2 \ldots d\tau_N$$

$\gamma$ is a square Hermitian matrix that in diagonal form can be represented as:

$$\gamma(1' \mid 1) = \sum_k \eta_k^* \eta_k n_k$$

here set $\{\eta\}$ is called natural spin-orbital according to Löwden[37]

The on-site Coulomb repulsion was varied over a range of values, including U = -10, -5, -2, -1, 0, 1, 2, 5, and 10 Hartree, to explore the behavior of the system under different interaction strengths.

In addition to studying the Hubbard model in isolation, the method was also applied to real-world systems. For example, the behavior of a water molecule was investigated using the 6-31G basis set, which is commonly used in quantum chemistry calculations. The properties of chains and rectangular and linear configurations of two $H_2$ molecules were also explored, using the ano-rcc-vdz basis set, which is a high-quality basis set that has been extensively tested and validated. The distance between atoms of each $H_2$ molecule was set up to 1 a.u.

Overall, the procedure for obtaining CI coefficient looks like this:
1. Generate 1 and 2 electron integrals for the system of interest
2. Compute reduced density matrices (RDMs) in spatial basis
3. Compute the NO basis by finding the eigenvectors of 1-body RDM matrix
4. Calculate new 1 and 2 electron integrals in natural orbital basis:

$$(a|b) = \sum_i \sum_j C_i^a (i|j) C_b^j$$

$$(ab \mid cd) = \sum_i \sum_j \sum_k \sum_l C_a^i C_b^j (ij \mid kl) C_c^k C_d^l$$

Where $C_n^k$ – n component of the $\eta_k$
5. Calculate energy and CI coefficients using FCI algorithm for the updated 1 and 2 electron integrals.

## Results and discussions

In order to prove that prosed ansatz exists, we performed the CI coefficients decomposition using a different order of approximation, so that:

$$\left(c_{l_{pred}}^{(k)}\right)^2 = \prod_{i_1 i_2 \ldots i_k} e^{-\omega_{i_1 i_2 \ldots i_k} n_{i_1} n_{i_2} \ldots n_{i_k}},$$

$$sign\left(c_{l_{pred}}^{(k)}\right) = \begin{cases} 1, if \; \dfrac{1}{1 + \exp\left(-\sum_i \widetilde{\omega}_i \; \widehat{X}_{il}^k\right)} > 0.5 \\ -1, otherwise \end{cases}$$

Here $\widehat{X}^k$ – feature vector for set of occupation numbers that is defined at equation (6). Some examples of predicted CI coefficients depending on different interatomic distances is shown at pictures below:

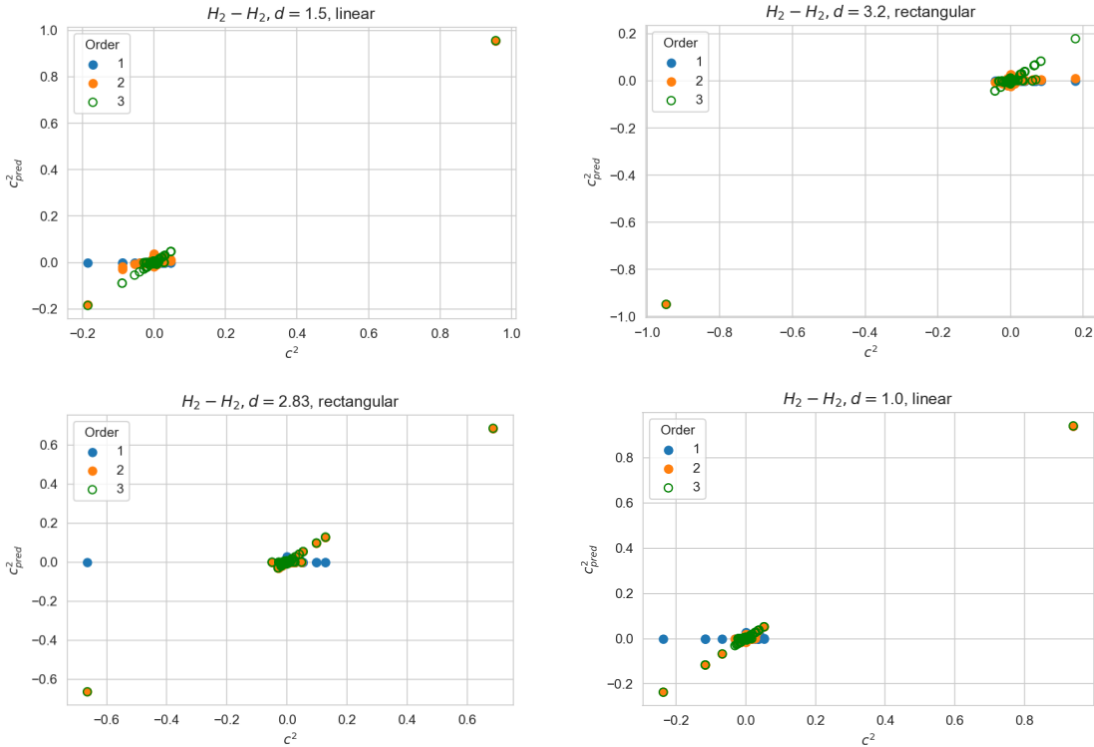

| order\system | linear, d=1.5 | rectangular, d=3.2 | rectangular, d=2.83 | linear, d=1.0 | rectangular, d=3.2; updated |
|---|---|---|---|---|---|
| 1 | (0.92, 0.91) | (0.9, 0.9) | (0.47, 0.47) | (0.89, 0.89) | (0.9, 0.9) |
| 2 | (0.96, 0.94) | (0.91, 0.89) | (0.97, 0.97) | (0.99, 0.99) | (0.97, 0.97) |
| 3 | (0.99, 0.99) | (0.98, 0.96) | (0.98, 0.98) | (0.99, 0.99) | (0.99, 0.99) |

*Table 1. Overlap and R^2 between the predicted and real wavefunctions for selected systems of 4 hydrogen atoms*

The results indicate that the method converges very quickly, with the second order approximation already providing an overlap between the predicted and true wave functions of >0.91, and $R^2$ >0.89. This is a significant improvement over the first order approximation, which had a lower overlap and $R^2$ coefficient, highlighting the importance of using higher order approximations for accurate predictions.

As expected, each new order of approximation was found to increase both the overlap and $R^2$ coefficient between the predicted and true wave functions, except for the system of rectangular configurated two $H_2$ molecules with spacing 3.2 a.u between the first and third atoms. In this case, the higher order approximations did not lead to a significant improvement in accuracy.

However, we were able to overcome this limitation by applying the same minimization algorithm with a larger step size at the stage of minimizing equation (4). This resulted in a significant improvement in the accuracy of the second order approximation, demonstrating the importance of carefully tuning the parameters of computational methods to achieve optimal results. Because the changes in the step size was applied at the second step of minimization algorithm that involves nonconvex minimization, we can conclude that algorithm just trapped at the local minimum.

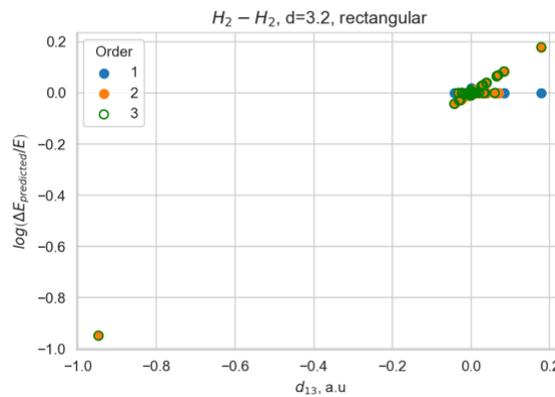

After the $c_{l_{pred}}^{(k)}$ obtained, we can calculate energy of the system using the predicted CI coefficients as:

$$E_{approx} = \sum_{ij} \rho_{ij}^{(1)} h_{ij} + \frac{1}{2} \sum_{ijkl} \rho_{ijkl}^{(2)} V_{ijkl},$$

Where $\rho^{(1)}$ – one electron RDM, $\rho^{(2)}$ —two electron density matrices, $h$ – one electron integral, $V$ – two electron integral.

Error in true ground state energy and approximated energy $E_{approx}$ that is calculated as $\log\left[abs\left(\frac{E_{true}-E_{approx}}{E_{true}}\right)\right]$ for the different h4 configurations:

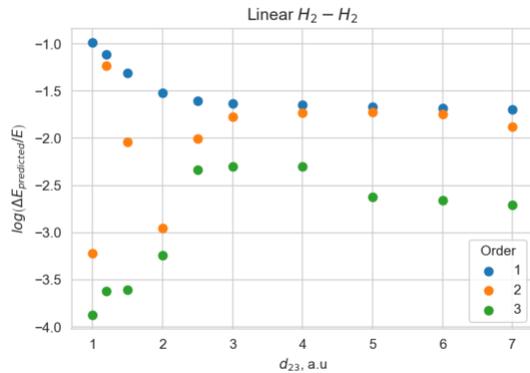
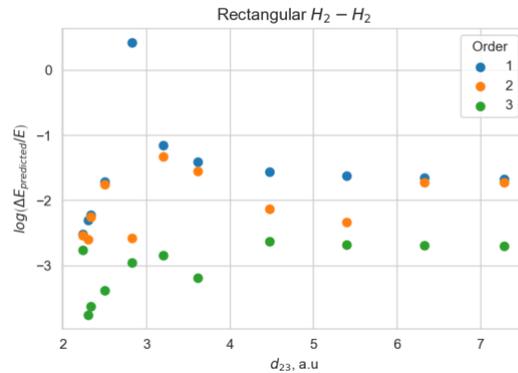

The results indicate that while the first order approximation can provide reasonably accurate results for many systems, the accuracy drops off for highly correlated systems. Specifically, the rectangular configuration of H4 system with the $d_{13} = 2.83$ a.u. proved to be a particularly challenging system for the first order approximation, with a relative logarithmic error of ~1.5. However, the second order approximation was found to be highly accurate, with a logarithmic relative error dropping below -2.5 for many systems, indicating an error of less than 5%. This is a significant improvement over the first order approximation and highlights the importance of using higher order approximations for strongly correlated systems.

Interestingly, the first order approximation was found to require only 4% of the initial set of CI coefficients, yet still produced reasonably accurate results with a relative error of approximately 20%. On the other hand, the second order approximation required a larger proportion of the initial set of CI coefficients, approximately 30%, but was able to achieve a much higher level of accuracy with a relative error of approximately 13%.

In order to test the limits of the proposed ansatz, we generated the systems described by Hubbard model with different repulsion potential U on the site.
Some of the coefficient regressions are shown at the pictures below.

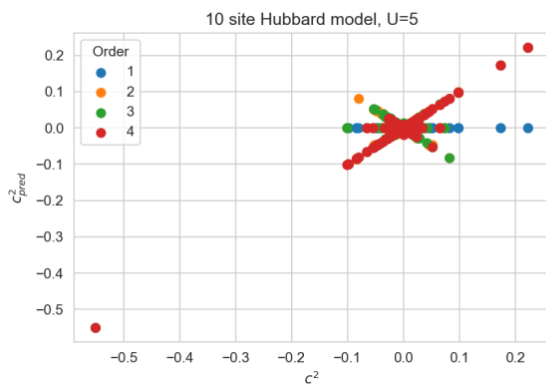
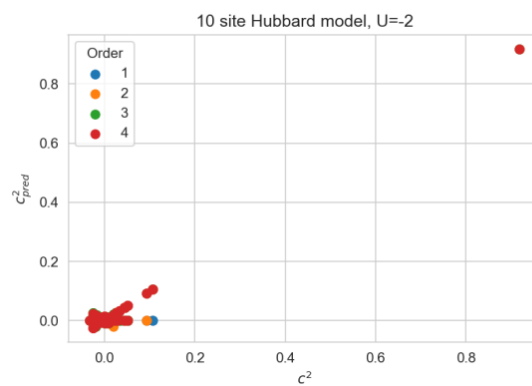

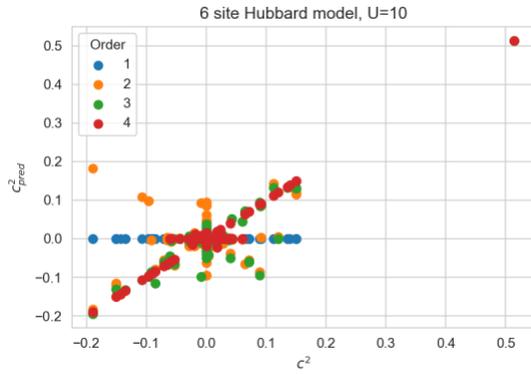 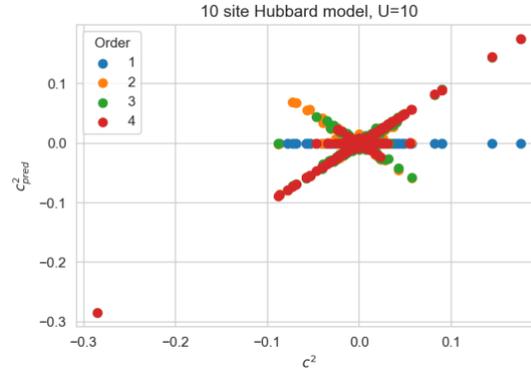

As expected, the bigger the absolute value of the repulsion potential, the slower ansatz is converging. Tables for the overlap and $R^2$ values are shown at tables 2,3:

| order\U | -10 | -5 | -2 | -1 | 0 | 1 | 2 | 5 | 10 |
|---|---|---|---|---|---|---|---|---|---|
| 0 | (0.08, 0.08) | (0.3, 0.3) | (0.84, 0.84) | (0.96, 0.96) | (1.0, 1.0) | (0.96, 0.96) | (0.84, 0.84) | (0.3, 0.3) | (0.08, 0.24) |
| 1 | (0.21, 0.17) | (0.43, 0.4) | (0.88, 0.87) | (0.97, 0.97) | (1.0, 1.0) | (0.98, 0.98) | (0.9, 0.89) | (0.54, 0.38) | (0.39, 1.34) |
| 2 | (0.36, 0.28) | (0.58, 0.56) | (0.9, 0.89) | (0.97, 0.97) | (1.0, 1.0) | (0.98, 0.98) | (0.92, 0.89) | (0.62, 0.51) | (0.52, 1.64) |
| 3 | (0.55, 0.4) | (0.67, 0.65) | (0.94, 0.92) | (0.98, 0.98) | (1.0, 1.0) | (0.98, 0.98) | (0.93, 0.92) | (0.74, 0.68) | (0.63, 1.91) |

Table 2. Overlap and $R^2$ between predicted and real wave function of 10 site Hubbard model with different repulsion terms

| order\U | -10 | -5 | -2 | -1 | 0 | 1 | 2 | 5 | 10 |
|---|---|---|---|---|---|---|---|---|---|
| 1 | (0.26, 0.25) | (0.57, 0.56) | (0.92, 0.92) | (0.98, 0.98) | (1.0, 1.0) | (0.98, 0.98) | (0.92, 0.92) | (0.57, 0.57) | (0.26, 0.26) |
| 2 | (0.57, 0.21) | (0.75, 0.75) | (0.96, 0.96) | (0.99, 0.99) | (1.0, 1.0) | (0.99, 0.99) | (0.97, 0.97) | (0.78, 0.6) | (0.57, 0.19) |
| 3 | (0.74, 0.53) | (0.89, 0.87) | (0.98, 0.98) | (0.99, 0.99) | (1.0, 1.0) | (1.0, 0.99) | (0.99, 0.98) | (0.91, 0.84) | (0.83, 0.7) |
| 4 | (0.91, 0.91) | (0.98, 0.95) | (1.0, 1.0) | (1.0, 1.0) | (1.0, 1.0) | (1.0, 1.0) | (1.0, 0.99) | (0.97, 0.95) | (0.93, 0.92) |
| 5 | (0.99, 0.98) | (0.99, 0.99) | (1.0, 0.99) | (1.0, 1.0) | (1.0, 1.0) | (1.0, 1.0) | (1.0, 1.0) | (1.0, 0.99) | (0.99, 0.99) |

Table 3. Overlap and $R^2$ between predicted and real wave function of 6 site Hubbard model with different repulsion terms

On the pictures below are shown the dependencies of the logarithm or the relative error from the different U for the 6 and 10 site Hubbard model.

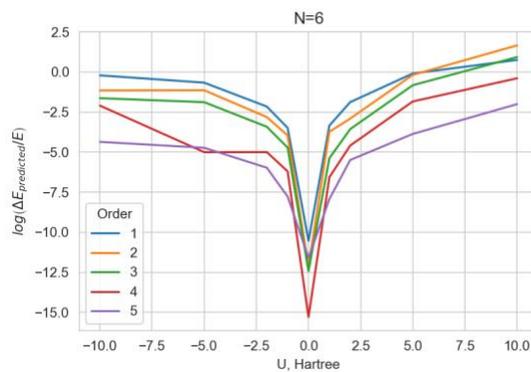 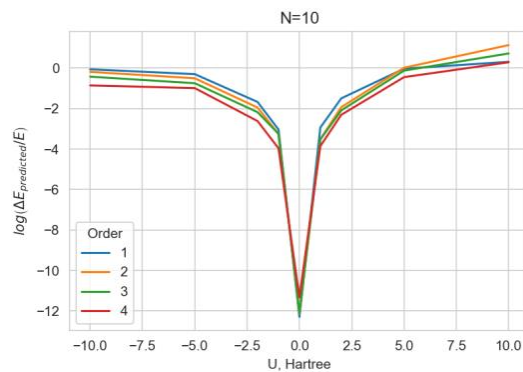

As expected, the model requires more parameters to achieve a better description of the energy as the level of correlations increases. Interestingly, when the on-site interaction is equal to 10 Hartree, the obtained energy values for both systems have error greater than 100%, indicating that we need even higher orders of approximation.

Additionally, it is important to note that the relative error is not equal for the system of 6 and 10 sites for the same U, even though the method is stated to be size consistent. It is happening because the energy was not minimized directly. Instead, the last step of minimization involved minimizing the squared difference between predicted and real CI coefficients of the systems. While this approach can be effective in achieving accurate predictions of the energy, it does not necessarily imply that the energy is minimized for a proposed ansatz.

As a result, future work could focus on the direct minimization of energy, which would not only prove the convergence of the proposed ansatz, but also demonstrate its size consistency. This would provide valuable insights into the accuracy and reliability of the model and could help to improve its effectiveness in modeling strongly correlated electronic systems.

## Summary

This publication proposes a new hierarchical wave function ansatz for theoretical and computational chemistry that combines insights from quantum mechanics and machine learning. The new wave function is expressed as weighted product of exponents of occupation numbers. The authors demonstrate the effectiveness of their approach by applying it to a range of benchmark molecular and one-dimensional lattice systems and show that it outperforms traditional wave function methods in terms of speed and computational efficiency. In addition to the improved efficiency, the proposed wave function ansatz has the benefit of being a straightforward, two-step minimization procedure. The optimization involves minimizing a convex function, which ensures computational stability and avoids the problem of local minima.

Future work for this proposed method includes direct energy minimization of the wave function ansatz, which may further improve its accuracy and efficiency. The authors note that their approach is generalizable and could potentially be extended to other quantum mechanical problems beyond electronic structure calculations. Overall, the proposed wave function ansatz provides a promising avenue for future research and development in theoretical and computational chemistry, with potential applications in various fields.